\documentclass[12pt,a4paper]{article}
\usepackage[a4paper,
bindingoffset=0.2in,
left=1in,
right=1in,
top=1in,
bottom=1in,
footskip=.25in]{geometry}

\usepackage{authblk}
\usepackage[utf8]{inputenc} 
\usepackage[T1]{fontenc}    
\usepackage{hyperref}       
\usepackage{url}            
\usepackage{booktabs}       
\usepackage{amsfonts, amsmath, amssymb}       
\usepackage{microtype}      
\usepackage{graphicx}
\usepackage[english]{babel}
\usepackage{cite}
\graphicspath{{Pictures/}}

\title{Full-Waveform Modeling for Time-of-Flight Measurements based on Arrival Time of Photons}
\author[1,*]{Maximilian Fink}
\author[2]{Michael Schardt}
\author[1]{Valentin Baier}
\author[1]{Kun Wang}
\author[1]{Martin Jakobi}
\author[1]{Alexander W. Koch}
\affil[1]{\small Technical University of Munich, Department of Electrical and Computer Engineering, Institute for Measurement Systems and Sensor Technology, Munich 80333, Germany}
\affil[2]{\small Blickfeld GmbH, Munich 80339, Germany}

\date{Preprint}
\begin{document}
	\maketitle
	\begin{abstract}
	\noindent Modern LiDAR sensors find increasing use in safety-critical applications. Therefore, highly accurate modeling of the system's behavior under demanding environmental conditions is necessary. In this paper, we present a modular structure to accurately simulate the amplified raw detector signal of a direct time-of-flight LiDAR system for coaxial transmitter-receiver optics. Our model describes, a measurement system based on standard optical components and a detector able of converting single photons to an electrical signal. To verify the model's predictions, single-point measurements for targets of different reflectivity at defined distances were performed. Statistical analysis shows an R-squared value greater than 0.990 for simulated and measured signal amplitude levels. Noise modeling shows good accordance with the performed measurements for different target irradiance levels. The presented results have a guiding significance in the modeling of the complex signal processing chain of LiDAR systems, as it enables the prediction of key parameters of the system early in the development process. Hence, unnecessary costs by design flaws can be mitigated. The modular structure allows easy adaption for arbitrary LiDAR systems.
\end{abstract}

	\section{Introduction}
	\label{sec:introduction}
	Light Detection and Ranging (LiDAR) is one of the key technologies for the interaction of computational intelligence-driven devices with their environment. Because of its high resolution and accuracy, it has gained a lot of popularity in various fields of application \cite{Thakur.2016, Ghorpade.2017}, where environmental perception is needed.
	As it is an active measurement method, the scene needs to be illuminated by usually a laser source. Therefore, different illumination schemes such as flash, 1D, or 2D scanning exist \cite{Villa.2021}. For scanning systems, a beam-steering mechanism is often needed, where the transmission and the reception paths are aligned in a coaxial manner \cite{Pennecot.7232014}. The most prominent ones include spinning polygon mirrors \cite{Engberg.582018}, rotating prisms \cite{Duma.2021}, MEMS scanners \cite{Wang.2020}, and optical phased arrays \cite{Hsu.2021}. The coaxial arrangement requires a coupling mechanism, e.g., spatial coupling, using the polarization of out- and in-going rays or a beam splitter. Scattering at the coupling element, the scanning mechanism, and other optical surfaces like lenses can cause a strong zero pulse directly after the laser emission seen by the detector. This zero pulse can affect the system response strongly, as it can lead to high intensity on the detector, causing saturation. It falsifies estimations based on pulse-width and amplitude of the received signal.\\
	Range information is extracted from the measurement signal according to the well-known formula for the direct time-of-flight (dToF) method \cite{Fersch.2017}:
	\begin{equation}
		d = c_0 \cdot \frac{t_{TOF}}{2},
		\label{eq:ToF}
	\end{equation}
	where $d$ denotes the object's distance, $c_0$ the speed of light, and $t_{TOF}$ is the measured time-of-flight. To extract $t_{TOF}$ from the amplified detector signal, different techniques have emerged for the scanning and the flash illumination patterns. Therefore, an analog-to-digital converter (ADC) or a threshold comparator followed by a time-to-digital converter (TDC) \cite{Liu.2020} is usually used. In this paper, the focus is on the prior stages, including an optical model, a detector model, and an amplifier model. Modeling of this system chain from the emission of the laser pulse until the amplification of the detector signal allows for an estimation of the system's detection range and helps to evaluate the performance of different system concepts and signal processing approaches.\\
	Estimates on the ranging performance of LiDAR systems and modeling of the analog signal of typically used detectors have previously been investigated \cite{Fersch.2017, Pasquinelli.2020}. Our contribution unites optical, electro-optical, and electronic impacts to the amplified detector signal, simulated in the time domain.\\
	In the following, the reference measurement setup, which shall be modeled, is presented. Afterward, the modular structure of the model is described in detail, exhibiting the needed input parameters for each step in the simulation. Single point measurements recorded with the measurement setup are used to validate the model with respect to the amplitude of the received signal. To evaluate the quality of noise modeling, that can either be from detector inherent noise or from background irradiation, the baseline shift and standard deviation of the amplified detector signal are investigated. 
	
	\section{Measurement Setup}
\begin{figure}
		\centering
		\includegraphics[scale=1]{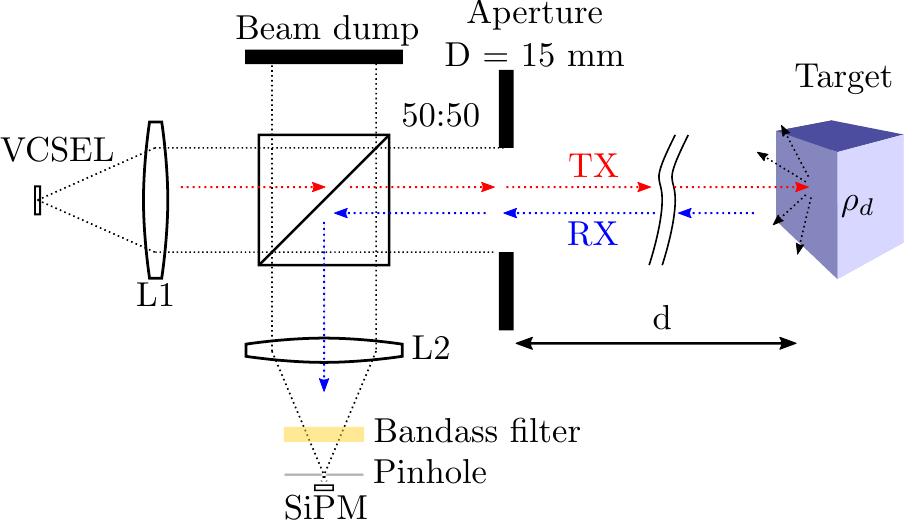}
		\caption{Measurement setup, with a VCSEL-array as pulsed light source and a SiPM as photodetector. Coupling of transmitting (TX) and receiving (RX) paths is achieved by using a beam splitter. One of the herewith generated two paths is blocked with a beam dump causing the zero pulse seen by the detector, while the other path is used for the distance measurement.}
		\label{fig:Setup}
	\end{figure}

	The presented measurement system is shown in Fig \ref{fig:Setup}. It consists of a vertical-cavity surface-emitting laser (VCSEL) array as pulsed light source and silicon photomultiplier (SiPM) as a detector. The laser beam needs to be transmitted (TX) along the same axis as the receiver's field of view to combine with a scanning mechanism, so a 50:50 beam splitter is used. Even if the beam splitter doesn't provide optimum coupling, the setup can be used for reference measurements and provide assessments of an optimized system. A single plano-convex lens (L1) was used to collimate the light from the squared VCSEL array with a divergence of 0.2° $\times$ 0.2°. On the receiver side, spectral filtering is achieved by using an optical bandpass filter with a spectral width of 10~nm. This helps to suppress  background irradiation, mainly caused by sunlight. Spatial filtering is implemented by a slit positioned in the focal plane of the receiving lens (L2) in front of the SiPM detector. This limits one dimension of the detector's FoV (field of view) in the target plane to the area that is illuminated by the laser. Therefore, the acceptance angle of the receiving (RX) path with the slit is 0.2° $\times$ 0.4°. If an exact overlap of the RX and TX cone is desired, a pinhole instead of a slit must be used.\\	
	An important parameter affecting the maximum achievable range of a ToF-LiDAR is the optical receiving aperture. For larger apertures, more of the photons reflected by the target can reach the detector. In scanning systems, this parameter is often limited by the scanning mechanism itself. For the evaluations in this paper, we used an aperture diameter of $D = 15$~mm. This value should be compatible with most scanning mechanisms.\\
	
\begin{figure}[h!]
		\centering
		\includegraphics[scale=1]{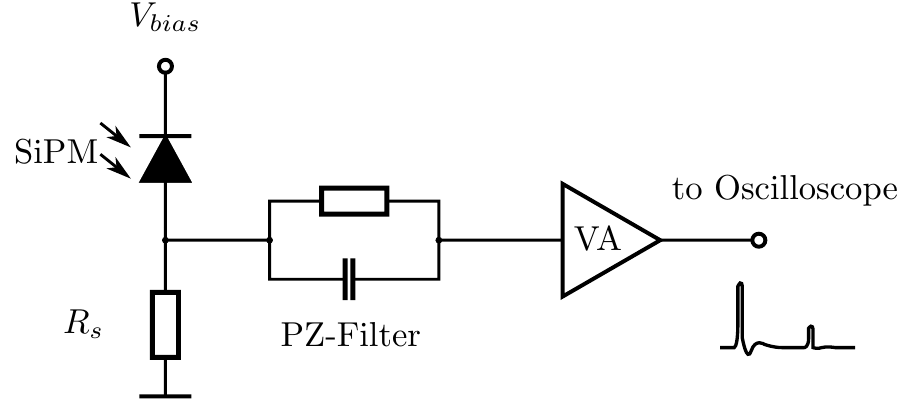}
		\caption{Schematic for read-out of the amplified detector signal with an oscilloscope. The current from the SiPM gets converted to a voltage via a shunt resistor $R_{s}$. This voltage is filtered by the pole-zero (PZ) filter \cite{Gola.2011}. The filtered voltage gets amplified by a voltage amplifier (VA). The amplified waveforms are captured with an oscilloscope.}
		\label{fig:Scope}
	\end{figure}

	The used VCSEL array consists of 12 individual VCSELs and has an emission wavelength of 808~nm. With a customized laser driver circuit, an output pulse width of 5~ns with an optical peak power of 2~W was measured. This leads to a pulse energy as low as 10~nJ. Despite conventional LiDAR systems using laser diodes, able to emit several hundred nanojoules in a single pulse, the low spectral drift, the symmetrical beam profile, and the continuous improvement of power conversion efficiency result in a high attractiveness of VCSELs for sensing applications like LiDAR \cite{Grabherr.2007, Wang.2015, Warren.2018, Maillard.2020}.\\
	Due to the very low pulse energy of the designed system, a highly sensitive photodetector is needed. Hence a SiPM is used as a detector. A SiPM is a parallel connection of individual SPADs (single-photon avalanche diodes) and allows detection thresholds down to the level of single photons \cite{Renker.2006}. The main advantage of SiPM over other conventionally used detectors in LiDAR, as APDs (avalanche photodiodes), is the low operating bias voltage and low-cost \cite{Adamo.2016}. The continuous enhancement of their sensitivity, also known as photon detection efficiency (PDE), for the near-infrared and shorter recovery times makes them increasingly attractive for use in such systems. \\ 
	As the signal obtained from the SiPM has a respectively slow fall-time, this can easily cause a pile-up of subsequent firing cells or cause a shift of the baseline if many cells fire simultaneously. Therefore we used the pole-zero compensation method presented in \cite{Gola.2011} to eliminate the slow fall-time of the SiPM signal and preserve its fast rise-time. This helps to prevent baseline shift due to pulse pile-up. The filtered signal is then amplified by a voltage amplifier (VA) circuit with a gain of 58 and a closed-loop bandwidth of 700~MHz. This high bandwidth helps to amplify the fast signals of the SiPM. To record the amplified signal, a high-speed oscilloscope was used. The schematic used for SiPM readout is shown in Fig. \ref{fig:Scope}. For the measurements at the SiPM, the reverse bias voltage $V_{bias}$ was applied according to the datasheet \cite{ONSemiconductor.2019}. 
	\\
	
	The zero pulse is caused by crosstalk from the emitter to the detector during emission of the laser pulse and is characteristic in coaxial systems. For this system, light can reach the detector in two ways: scattering at the beam splitter and diffuse reflection at the beam dump. The second path dominates, as half of the transmitted power emitted by the laser diode is deflected towards the beam dump. Even if only 0.01~\% of the incident energy on the beam dump is scattered back to the detector (ca. 2~$\times$~10\textsuperscript{10} photons), the zero-pulse's intensity is much higher than most reflections of objects in the measurement path. In other systems, this zero reflection can be caused by scattering at various optical surfaces like the scanning mechanism itself or the housing window. In the case of a SiPM as detector, the zero pulse will cause the majority of the microcells to fire, and therefore cause a peak in the observed detector signal. The released charge by the SiPM during this event will also cause the following amplifier circuit to leave its linear operation mode and go into saturation. The amplifiers overdrive recovery time should therefore be considered and kept as short as possible.\\
	
	\section{Model Structure}
\begin{figure}
		\centering
		\includegraphics[scale=1]{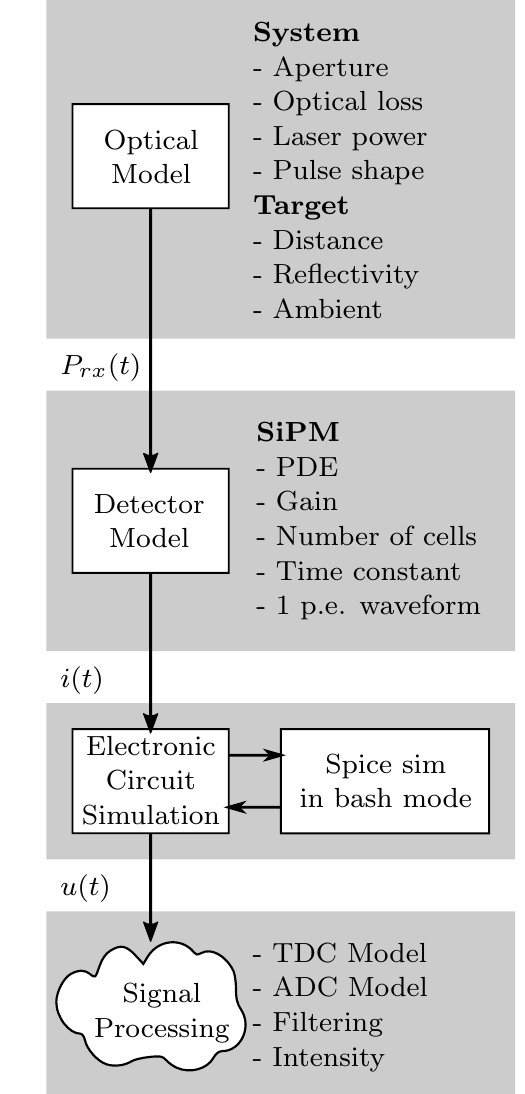}
		\caption{General Structure of the model. For each stage of the model, the necessary input parameters are listed. For the analog front-end model, a SPICE simulation was used in our case.}
		\label{fig:Model}
	\end{figure}

	To model the output waveform of the sample measurement system, the presented structure focuses on three essential parts of the system: the optical system, the detector, and the analog front-end electronics to amplify the detector signal. A block diagram of the proposed model structure and its input parameters are shown in Fig. \ref{fig:Model}. The simulated voltage waveform can be used to test further processing stages and algorithms which convert the analog signal to discrete voltage levels and then perform a peak detection. This paper focuses on the correct modeling of the first three stages of the model.   
	
	\subsection{Optical Model}
	The returned signal power $P_{rx, s}$ within the systems aperture $A_0$ when emitting the power $P_{tx}$ towards a Lambertian target with a diffuse reflectivity $\rho_d$ in a distance $d$ can be calculated according to \cite{Adams.2002}:
	\begin{equation}
		\label{eq:PLSR}
		P_{rx, s} = P_{tx} \cdot A_0 \cdot L_{s}(d) \cdot \frac{\rho_d \cdot \cos\Theta}{\pi}.
	\end{equation} 
	$\Theta$ describes the angle of incidence of the emitted laser beam on the object, and $L_{s}(d)$ describes an optical loss factor for the signal path. In the case of the beam splitter of the sample system, we define:
	\begin{equation}
		L_{s, ideal}(d) = \frac{1}{4d^2}.
	\end{equation}
	
\begin{figure}[h!]
		\centering
		\includegraphics[scale=1]{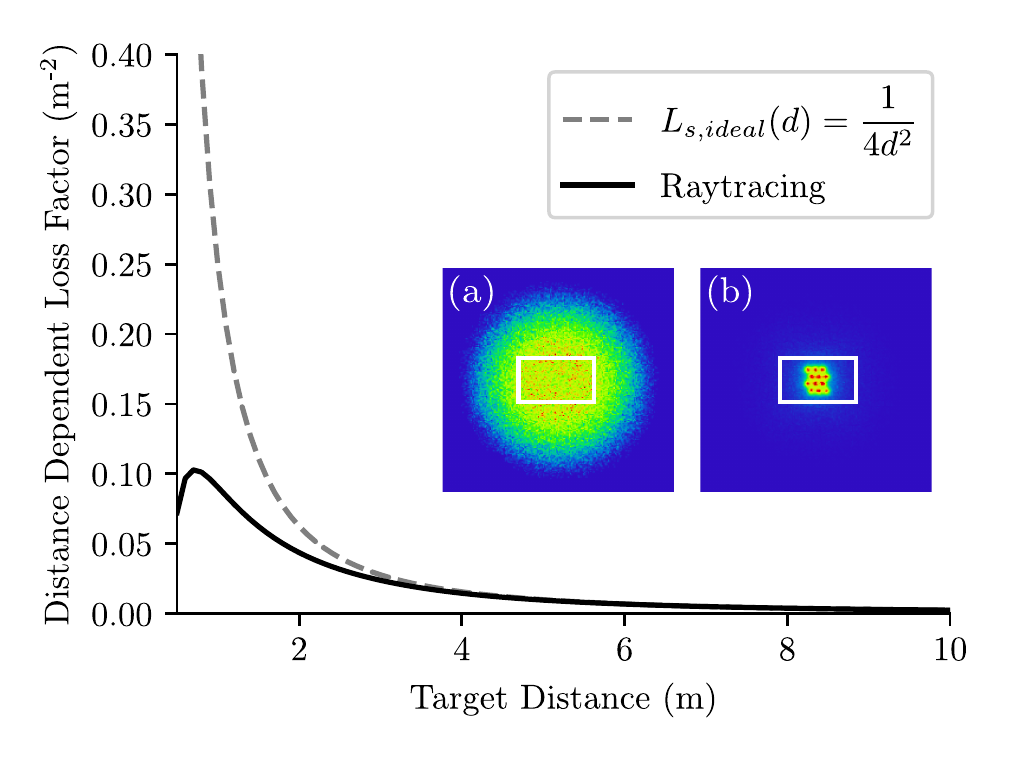}
		\caption{Calculated and simulated $L_{opt}(d)$. Especially for distances $d<$4 m, a deviation can be seen. This can be explained by looking at the simulated intensity distribution in the pinhole plane for a target in (a) 1 m and (b) 20 m. For very close targets, not all of the received power is passing the pinhole indicated by the white rectangle.}
		\label{fig:GeomLoss}
	\end{figure}

	In coaxial systems, the optical loss factor depends on the coupling mechanism used. Additionally to the general course for the defined loss factor, in coaxial systems, additional losses can occur. These can most likely be compared to the geometric loss in biaxial LiDAR systems, determined by the evolving overlap of the transmitting and receiving cone over the measurement distance \cite{VandeHey.2011}.
	An investigation of the intensity in the detector plane with the optical design software OpticStudio\textsuperscript{\textregistered} was carried out to find these additional losses. 
	Fig. \ref{fig:GeomLoss} shows the curve for the ideal case and the results from the raytracing simulation. A deviation from the ideal curve for targets closer than 4~m can be identified. An additional loss is visible when looking at the simulated intensity in the spatial filtering plane of the system. For close targets, the laser spot is out of focus. The returned power passes the pinhole only partially, as shown in Fig. \ref{fig:GeomLoss} (a). For increasing measurement distances, this effect decreases until all the received power can be passed through the pinhole towards the detector (Fig. \ref{fig:GeomLoss} (b)). For the latter shown measurements, this effect has only a slight impact. It is necessary to highlight that it can be of relevance for other system architectures, where such effects, caused by the coupling mechanism to get coaxial, can affect the system response up to its maximum measurable range.\\
	
	The returned noise power can be calculated by the detector's FoV, defined by the horizontal $\theta_h$ and vertical divergence $\theta_v$ in the target plane, as well as the optical power density $S$ in W/m\textsuperscript{2} caused by sunlight after bandpass filtering. For the sample system, light from external sources needs to pass the beam splitter only once, and therefore, the loss factor for noise sources like daylight is $L_{n}(d) = \frac{1}{2d^2}$.
	Because the area that the detector sees increases with distance $d$, this variable cancels out from the equation for the received power for external sources. For the given measurement system, the returned noise power can be calculated by:
	
	\begin{equation}
		P_{rx, n} = \frac{\theta_h \cdot \theta_v \cdot S \cdot A_0 \cdot \rho_d}{2\pi},
		\label{eq:PDL}
	\end{equation}  
	using the small-angle approximation. To produce a signal in the time domain, the maximum measurement time is divided into small intervals. In this work, a time resolution of 500~ps was used, as this was found to be a good trade of between accuracy of the predicted values and simulation speed. Given the time base, the energy for a time slice for the received laser light (Eq.~(\ref{eq:PLSR})) and ambient light (Eq.~(\ref{eq:PDL})) can be calculated and converted to an equivalent mean number of photons. For the distribution of the laser photons, the laser pulse shape is also considered. We, therefore, assume the laser pulse to have a truncated Gaussian shape and scale the number of expected photons according to its amplitude. For each time step, a single draw from a Poisson distribution with the calculated mean number of photons is performed, according to photon statistics \cite{Fox.2006}.
	Also, the photons by the internal reflection $N_{zero}$ causing the zero pulse are modeled in this step. As described earlier, the photon count for this is several orders of magnitude higher when compared to the real signal photon levels. The generated photon vector can then be fed to the detector model.
	
	\subsection{Detector Model}
	This part of the model reproduces the behavior of the detector, which converts impinging photons into electrical charge. Simulating SiPMs has been investigated by several previous works \cite{Pasquinelli.2020, Acerbi.2019}. Utilizing a SiPM as a detector in LiDAR systems has been shown for biaxial systems \cite{Adamo.2016}. As SiPMs mainly target applications where counting of single photons is necessary, their behavior for two subsequent pulses that can have a high intensity is not considered for most applications. As this is the exact case for our measurement system, the effects occurring have to be taken into account. Only recent investigations show their nonlinear response for this case \cite{Bretz.2020}. Also, their behavior under continuous light \cite{Nagai.2019} has to be considered for a LiDAR system.\\
	
	Our model assumes SiPM saturation by the internal reflection, meaning all microcells fire within the time of the laser emission.  After that, cells recover with the recharge time constant $\tau_r$ implying that trigger probability $PDE(t)$ and gain $G(t)$ of the microcells recover according to the overvoltage $V_{ov}$ with:
	\begin{equation}
		V_{ov}(t) = (V_{bias}-V_{brd}) \cdot \bigg(1-e^\frac{t-t_0}{\tau_r}\bigg),
		\label{eq:Recovery}
	\end{equation}
	whereas $t_0$ denotes the time where the emission of the laser ends, $V_{bias}$ refers to the applied reverse bias voltage, and $V_{brd}$ to the SiPMs breakdown voltage. \\
	Hence,  a simple model for $PDE(t)$ and $G(t)$ for the expected behavior after saturation considering specified PDE in the datasheet $PDE_0$ and expected gain $G_0$ is given by:
	\begin{align}
		PDE(t) &=  PDE_0 \cdot (1-e^\frac{t-t_0}{\tau_r}),\label{eq:PDE}\\	
		G(t) &= G_0 \cdot (1-e^\frac{t-t_0}{\tau_r}). 
		\label{eq:Gain}
	\end{align}
	
\begin{figure}[h!]
		\centering
		\includegraphics[scale=1]{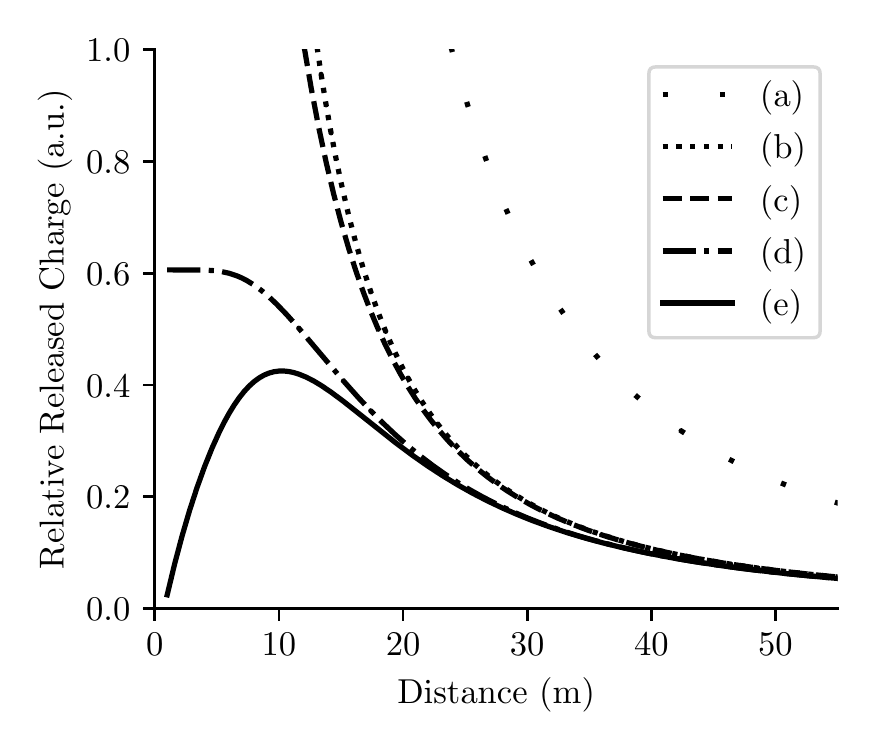}
		\caption{Released Charge for a SiPM detector in a coaxial LiDAR system if we assume a zero pulse is causing all microcells to trigger. (a) ideal SiPM, (b) SiPM with $PDE<1$, (c) additionally considers time-dependent PDE according to equation (\ref{eq:PDE}), (d) additionally considers a limited number of cells according to equation (\ref{eq:cells}), (e) additionally considers a time-dependent gain due to the recharge process expressed by equation (\ref{eq:Gain}).}
		\label{fig:Effects}
	\end{figure}

	With the knowledge of the behavior of the PDE and the gain, an equivalent number of fired cells can be calculated. The impact of  the following individual described effects on the detector's response can be seen in Fig. \ref{fig:Effects}. Using the predicted incident photons on the detector at each time step $t$, calculated by the optical model, the following steps are performed in the detector model:
	\begin{enumerate}
		\item Calculate theoretically detected photons $N_{det}(t)$ for $PDE(t)$ in Eq. (\ref{eq:PDE}) and the number of impinging photons $N_{ph}(t)$ for each time step $t$:
		\begin{equation}
			N_{det}(t) = PDE(t) \cdot N_{ph}(t).
		\end{equation}
		\item Determine the number of fired cells $N_{fired}(t)$, according to the total existent number of cells $N_{tot}$ in the device: 
		\begin{equation}
			N_{fired}(t) = N_{tot} \cdot \bigg(1-e^{-\frac{N_{det}(t)}{N_{tot}}}\bigg).
			\label{eq:cells}
		\end{equation}
		\item Identify the equivalent number of fired cells $N_{eq}(t)$ considering the released charge with the number of fired cells $N_{fired}(t)$ and the current gain $G(t)$:
		\begin{equation}
			N_{eq}(t) = \frac{G(t)}{G_0}N_{fired}(t).
		\end{equation} 
	\end{enumerate}
	
	The hereby estimated equivalent number of fired cells $N_{eq}(t)$ can be used to determine the current waveform $i(t)$ in the time domain. In our case, interpolation with the measured single-photon waveform is performed, but it can also be used as input for SiPM spice models like \cite{Acerbi.2019}. \\
	Besides the detection of the signal and ambient photons, intrinsic detector noise such as the dark count rate (DCR), crosstalk (XT), and afterpulsing (AP) \cite{Piemonte.2019} are also added in this step.
	The DCR is calculated independently from impinging photons to the detector. For each time step with the width $\Delta t$, the number of dark counts is determined by the Poisson distribution with the mean $DCR \times \Delta t$. Crosstalk is considered using the crosstalk probability $P_{XT}$ and fired cells in the actual time step $N_{fired}(t)$. For afterpulsing events, the corresponding probability $P_{AP}$ is multiplied with the number of fired cells from the previous time step $N_{fired}(t-\Delta t)$. 
	
	\begin{figure}[h!]
		\centering
		\includegraphics[scale=1]{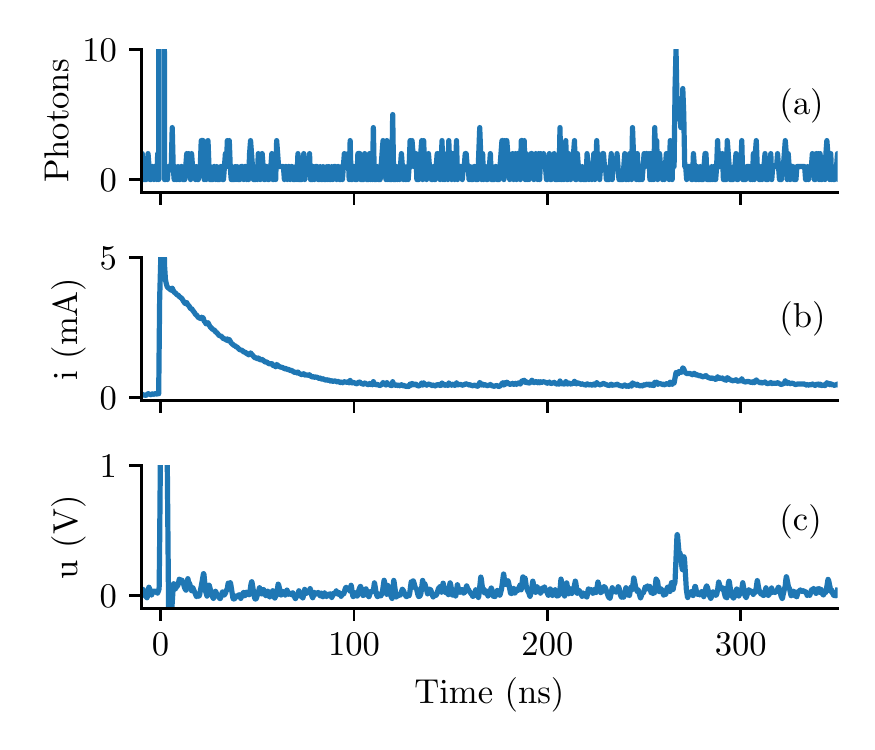}
		\caption{Output generated in each step of the model. A single waveform for the exemplary parameters $\rho_d\text{ = 0.5}$, $d$ = 40 m, and a daylight level of 40 klx is shown. A time resolution of 500~ps is used. (a) Incident photons to the detector generated by the optical model. (b) Current waveform based on incident photons and intrinsic noise effects. (c) A converted, filtered, and amplified LiDAR signal is generated by an electrical model of the used analog-front-end circuit.}
		\label{fig:StepsModel}
	\end{figure}
	
	\subsection{Analog Front End Model}
	The generated current waveform $i(t)$ is used as input for the model of the analog front-end electronics. Therefore, we used an implementation of the used amplifier circuit in an arbitrary electronic circuit simulator. Otherwise, a transfer function of the circuit can be used. The SPICE model, though, provides modeling of saturation behavior, by modeling e.g. overdrive recovery of operational amplifiers, and is, therefore, more accurate than a simple transfer function. It also allows investigation of environmental effects on the system, such as temperature, at a component level. Typical amplifier circuits for SiPMs are either a transimpedance amplifier or a shunt-resistor in combination with a voltage amplifier \cite{Calo.2019}. As mentioned in the chapter before, we did use the analog filter presented in \cite{Gola.2011} to get rid of the slow tail of the SiPM signal in combination with an operational amplifier-based voltage amplifier. \\
	
	The presented modular structure of the model allows changes to be implemented on every part of the system individually. Figure \ref{fig:StepsModel} shows the corresponding waveforms for the three presented steps for an exemplary set of parameters. 
	
	\section{Measurements and Simulation Results}
	To verify our model, we conducted single-point measurements with the presented setup. As a beam dump a piece of black foam was used, found to reduce reflection towards the detector. Measurements have been accomplished on two calibrated targets with Lambertian reflectivity of 10~\% and 50~\%, starting at a distance of 5~m up to a distance of 40~m. For each measurement point, 1000 waveforms have been acquired, with an oscilloscope having a bandwidth of 700~MHz. The waveforms have been captured at the output of the amplification circuit, as shown in Fig. \ref{fig:Scope}.
	
	\subsection{Evaluation of Signal Peak Amplitude}
\begin{figure}
		\centering
		\includegraphics[scale=1]{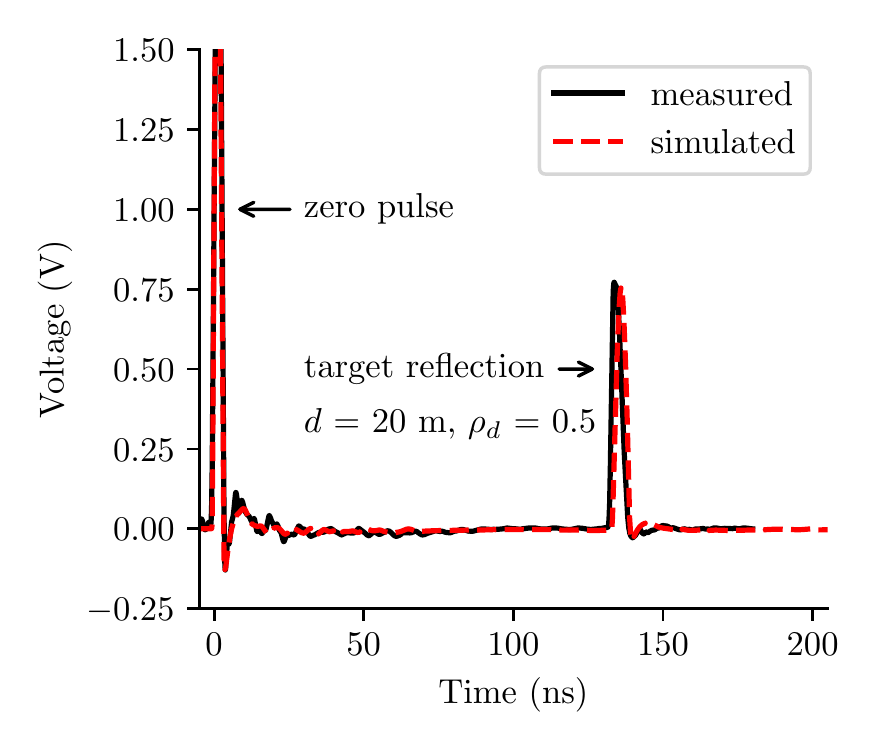}
		\caption{Average measured waveform and corresponding simulated waveform for a target in a distance of d = 20 m and a diffuse reflectivity of $\rho_d$ = 0.5. At the internal reflection, the amplifier exceeds its linear region and hard clipping occurs.}
		\label{fig:Waves}
	\end{figure}

	Figure \ref{fig:Waves} shows a direct comparison between an averaged waveform for the simulation and the measurement for a target with specific reflectivity of $\rho_d$ = 0.5 in a distance of $d$ = 20 m. The first peak in the amplified detector signal is caused by the internal reflection. Due to the high intensity of the internal reflection, the linear range of the amplifying circuit is exceeded, and clipping of the signal occurs due to saturation at an amplitude of about 2 V. Therefore, it is important to use an amplifier with a fast overdrive recovery to minimize the width of the internal reflection peak (as it will be impossible to detect targets closer than the duration of this zero pulse). After this, a short ringing effect can be noticed. The measurement wave also shows some disturbances caused by electromagnetic interference from the laser circuit. The expected peak in the signal of the target reflection can be found at 133~ns, according to equation (\ref{eq:ToF}). A good agreement for the measured and simulated waveform can be achieved. 
	
	\begin{figure}
		\centering
		\includegraphics[scale=1]{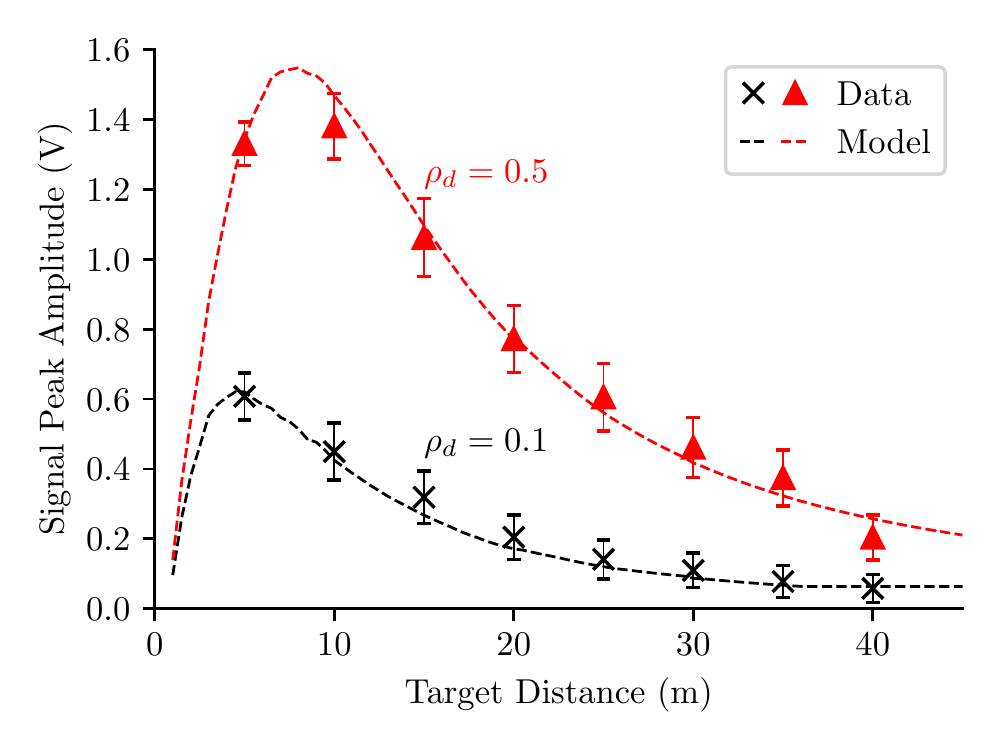}
		\caption{Measured and simulated amplitudes of the received signal over distance for two targets with different reflectivity. The standard deviation of the measured signal peak amplitudes is indicated.}
		\label{fig:Amplitudes}
	\end{figure}
\begin{figure}
		\centering
		\includegraphics[scale=1]{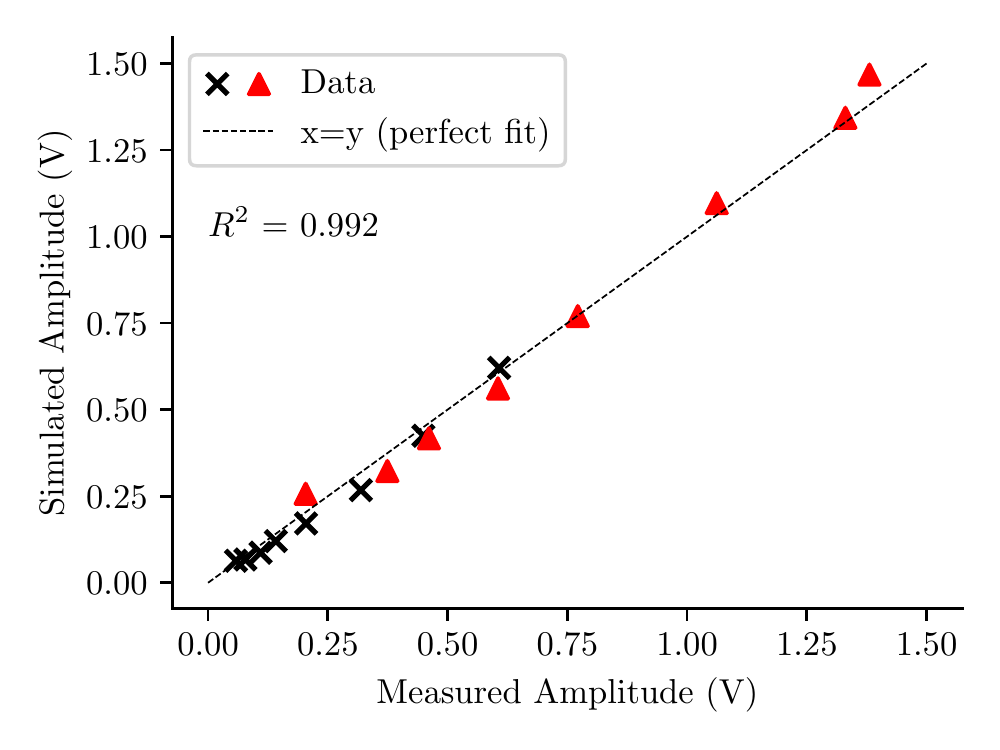}
		\caption{Mean of the measured signal amplitudes plotted against the predicted amplitudes of the model. Linear regression gives a high $R^2$ value.}
		\label{fig:ScatterPlot}
	\end{figure}

	For comparison with the model, the average height of the measured signal peaks caused by the target reflection was extracted and plotted in Fig. \ref{fig:Amplitudes}. The signal peak height can be approximated well by the presented model. If compared to the modeled detector effects depicted in Fig. \ref{fig:Effects}, it can be seen that this behavior is dominating the response of the system. Besides the mean value of the signal peak amplitude caused by the target reflection, Fig. \ref{fig:Amplitudes} also indicates the standard deviation of the investigated value for the 1000 recorded waves. For target distances greater than 15 m, effects caused by detector saturation do not affect the system response strongly anymore. For this area, the standard deviation decreases for lower received photon counts, which corresponds to the Poisson nature of photon detection.\\
	If measured mean signal peaks of the target reflection are directly compared against the predicted values by the model, a high linear regression coefficient value of $R^2$ = 0.992 is exhibited (see Fig. \ref{fig:ScatterPlot}). This indicates the validity of the modeled optical and detector effects in the presented model. 
	
	\subsection{Evaluation of Noise Modeling}
\begin{figure}
		\centering
		\includegraphics[scale=1]{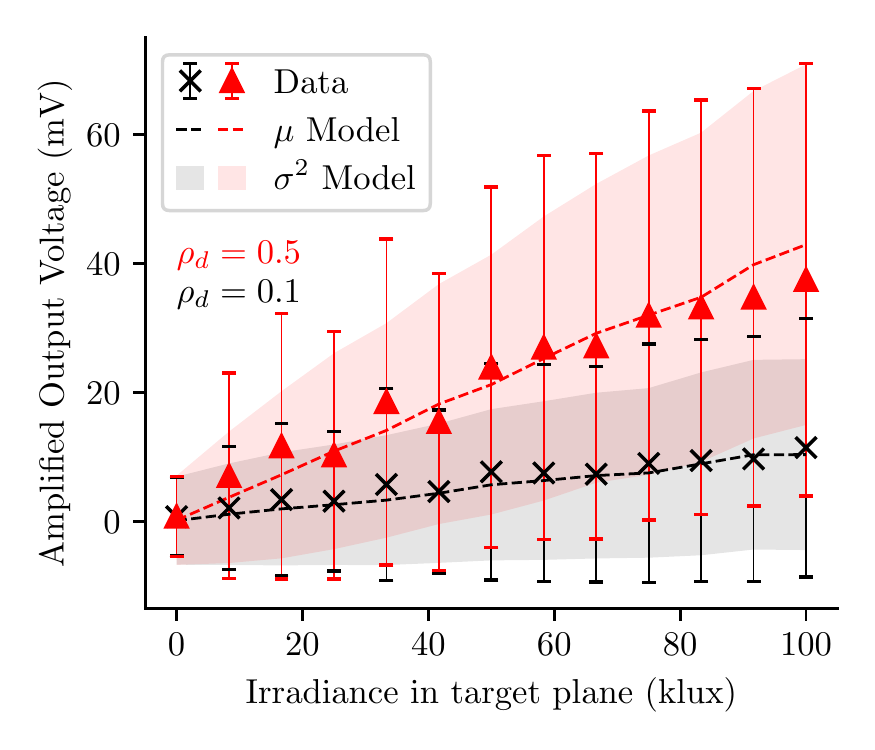}
		\caption{Measured and simulated offset voltage and standard deviation of the amplified detector signal for different irradiances in the target plane.}
		\label{fig:NoiseEval}
	\end{figure}

	To evaluate the noise modeling, we investigated the baseline shift and standard deviation of the amplifier output \cite{Nagai.2019}. Fig. \ref{fig:NoiseEval} shows the results for the measured data and the predictions by the model for two different targets. If there is a constant irradiance in the target plane, then the power reaching the detector is independent of the target distance described by Eq. (\ref{eq:PDL}). For LiDAR applications, the ambient lightning situation is usually expressed in klx. By measuring the power density with our bandpass filter in the target plane, the corresponding values can be calculated by using a conversion factor taking into account the spectral irradiance of the sun and the luminosity function.\\
	When comparing the measured means to the simulated means of the signals baseline, a good fit with a maximum deviation of ca. 2 mV for the 10~\% target, and ca. 4 mV for the 50~\% target can be investigated. Higher deviations of the standard deviations can be seen with a maximum error of 5~mV respectively 8.5 mV. This deviation can possibly be derived from self-heating effects increasing intrinsic detector noise, or also non-exact modeled amplifier bandwidth, as short peaks, representing higher frequencies, would be damped.

	\section{Conclusion \& Outlook}
	This paper introduced an approach to simulate the amplified detector signal of a dToF-LiDAR system. Relevant impacts arising from the optical, electro-optical, and electronic building blocks for a LiDAR system have been modeled. The model has been proved with a LiDAR system for point measurements based on standard optical components and low-cost semiconductors. We showed that the waveform of interest can be predicted accurately with knowledge of the parameters of the system. Due to its modular structure, the model can be adapted to arbitrary dTOF designs.\\
	For the optical system, knowledge about the optical aperture, the laser energy, and laser pulse shape, as well as the divergence of the transmitting and receiving cone, have to be known. For the hereby shown system with a SiPM as a detector, the parameters PDE, gain, dark count rate (DCR), crosstalk probability, and afterpulsing probability must be known. For the electronic amplifier, full knowledge of the used circuitry and component values should be acquainted in order to create an accurate SPICE model. Otherwise, a transfer function can also be used.\\ 
	Future work can be carried out by including environmental effects such as temperature, on the amplified detector signal. A temperature sweep on the component level could show system performance for different ambient temperatures. Further, different demanding weather conditions like rain, snowfall, or fog could be integrated into the model to predict the performance of the LiDAR system in harsh environments. The processing of the hereby simulated signal, such as sampling by an ADC or thresholding with comparators, and the following algorithms to extract the ranging information could be simulated to investigate the difference in varying implementations and help to choose the best performing one. The presented model can also be used to simulate LiDAR data on point cloud level if further components such as a scanning mechanism and housing windows are used and knowledge about the scan pattern is available. This will lead to new possibilities in the evaluation of the ongoing development of LiDAR systems, as system performance can already be predicted in the design phase.

	\section*{Acknowledgement}
	This work was supported by the BMWi of Germany under ZIM (Central Innovation Programme for small and medium-sized enterprises (SMEs)) [Grant numbers ZF4304604AB9 and ZF4615702AB9].
	
	\bibliographystyle{unsrt}
	\bibliography{Literature}
	
\end{document}